\documentclass[prb,twocolumn,showpacs,amsmath,amssymb]{revtex4}

\usepackage{graphicx}
\usepackage{amssymb,amsmath}
\usepackage{dcolumn}
\usepackage{bm}
\usepackage{color}

\begin{document}

\title{Acoustic control of the lasing threshold in QDs ensemble coupled to an optical microcavity}

\author{D.V. Vishnevsky}
\affiliation{LASMEA, Nanostructure and Nanophotonics group, Clermont Universit\'{e}, Universit\'{e} Blaise Pascal, CNRS, 63177 Aubi\`{e}re Cedex France}

\author{N.A. Gippius}
\affiliation{A. M. Prokhorov General Physics Institute, RAS, Vavilova Street 38, Moscow 119991, Russia}
\affiliation{LASMEA, Nanostructure and Nanophotonics group,
Clermont Universit\'{e}, Universit\'{e} Blaise Pascal, CNRS, 63177 Aubi\`{e}re Cedex France}

\begin{abstract}
We propose a theoretical model which describes the coupling between quantum dots ensemble and optical microcavity. By this model we simulate the interaction of the system with the strain pulse which can strongly modify the lasing of QDs.
\end{abstract}

\maketitle

\section{Introduction}
Nanoscale semiconductor heterostructures which are under intensive studies for more than 40 years\cite{Alferov} and particularly quantum dots (QDs) play an important role in modern physics. Electrons and holes in such structures are confined in all directions and their states become quantized. The interaction of the light with QDs can be strongly modified in case the QDs are put into the semiconductor microcavity (MC).
In such structures electromagnetic wave is confined between two mirrors, what strongly modifies the light behavior for example stop-bands and standing electromagnetic waves can be observed\cite{Microcavities, Weisbuch_PRL_69}.

Such systems are good objects for investigation of light-matter coupling. Depending on the interaction strength between QDs and MC two regimes are possible. First - weak-coupling regime shows slight modification of the photon dispersions. Photons and QDs excitons can be considered as different particles in this case. For large enough density of the quantum dots and small their inhomogeneous broadening one can expect also the  called strong-coupling regime. It's characterized by strong interaction constant and it gives strong modification of the particles behavior. However in this paper we will consider the week coupling regime.

Introducing to the system the acoustic field\cite{Akimov_PRL_97,Scherbakov_PRL_99,Berstermann_PRB_80}  one can observe significant effects, like acoustically driven amplifying of the lasing \cite{Bruggemann}. Acousto-optical interaction is based on a fact, that width of a band gap in semiconductors depends on lattice constant which value can be driven by acoustic pulse. And so the spectral possition of quantum dots could be modified by acoustic vibrations. Introducing a strain pulse to the system of quantum dots coupled to a microcavity during lasing one can involve more dots in the process of photoluminescence and increase the lasing power.

In our work we apply the theoretical model which describes an ensemble of quantum dots in optical microcavity. Here we consider quantum dots as a number of two-level systems, so the evolution of QDs population can be described by von Neumann equation. For the electromagnetic field of the microcavity we can use a resonant mode approximation \cite{Gippius_OE_18,Gippius_EPL_67,Gippius_JOP_16},  and get a system of two coupled equations which gives us a behavior of the system.

In first part of the paper we describe in more details the model we propose and its possibilities and limitations. We show several results that could be obtained with this model. And after that we introduce the acousto-optic interaction to our system to describe theoretically experimental results\cite{Bruggemann}. In the final part of this work we discuss a proposal of implementation of the surface acoustic waves to this system and describe effects of dynamical lasing patterns formation.

\section{Theoretical model}
\subsection{Quantum dots}

We use the density matrix formalism in a scalar approximation\cite{Krizhanovskii_ssc_01,Kulakovskii_pss_02} where each dot  can be described by its own 2x2 density matrix $\rho$. The dynamics of the system is given by von Neumannn equation:
\begin{equation}
i\hbar \dot \rho  = [\hat H\rho ]
\label{Neumann}
\end{equation}
Diagonal elements of $\rho$ give the probabilities to find quantum dot in ground or excited state and non-diagonal ones give the correlations between the ground and excited states, that are responsible for the magnitude of the quantum dot polarization. In fact Hamiltonian and density matrix depend on in-plane dot position. However for the electromagnetic wave propagating in $z$-direction (that is the subject of our study) field is homogeneous through $x$ and $y$ and all the dots are in the same field. Thus we can consider the quantum dots ensemble to be homogeneous in the plane. Because we neglect all the interactions between the dots except the one from the electromagnetic field the dynamics of each dot depends only on the local electromagnetic field and the resonant dot transition energy. The distribution of the dots over the spectrum is accounted for by the spectral density function of the dots $n_{x}(E^{i})$,  with the dot optical transition energy $E^{i}$.

Next approximation of our model is that we consider only two types of interaction with the quantum dots: first is the interaction of quantum dot with resonant electromagnetic field in the cavity that is explicitly put in Hamiltonian.
The second is external pumping and relaxation of quantum dots, that is introduced as an additional phenomenological terms in the equation for the density matrix of $i$-th type of quantum dots:
\begin{equation}
i\hbar \dot \rho ^i  = [\hat H^i \rho ^i ] + P^i
\label{Neumannk}
\end{equation}
The Hamiltonian of $i$-th type of quantum dot with resonant energy $E^i$ can be written in the following form:
\begin{equation}
\hat H = \left( {\begin{array}{*{20}c}
   0 & {(i{\cal E }_0 d^i )^* }  \\
   {i{\cal E }_0 d^i } & {E^i }  \\
\end{array}} \right)
\label{Hamilton}
\end{equation}
$d^i$ - it is the dipole moment matrix element between the ground and excited states of $i$-th quantum dot, ${\cal E }_0$ - amplitude of the cavity electromagnetic field on quantum dots layer, $E^i$ - energy of the first exciton state in the QD.
The dipole matrix element of the dot can be expressed via martix element of the interband currents of QD of $i$-type: $d^i=\frac{J_0^i}{\omega_0^i}$. Its value is defined only by quantum dot structure and we take it as a parameter in our model, the $\omega_0^i=E_i/\hbar$ is a frequency of QD optical transition.

The last term in Eq.(\ref{Neumannk}) responsible for external pumping and relaxations in QDs reads as:
\begin{equation}
 P^i = \left( {\begin{array}{*{20}c}
   { - (\rho _{22}^{st}  - \rho _{22}^i ){\gamma_1}} & {-\rho_{12}^i\gamma_2} \\
   {-\rho_{21}^i\gamma_2} & {  (\rho _{22}^{st}  - \rho _{22}^i ){\gamma_1}}  \\
\end{array}} \right)
\label{Pump}
\end{equation}
If one neglect the interaction with the cavity mode
the nonresonant pumping drives the system to some stationary state with $\rho _{22}^i=\rho _{22}^{st}$ within the characteristic relaxation time $1/\gamma_1$. As a first approximation we assume that $\rho _{22}^{st}$ is the same for all types of quantum dots. The non-diagonal elements relax to zero with characteristic time $1/\gamma_2$.
All these values are taken as parameters of the model. In order to  describe the spontaneous emission in QDs ensemble we add a white noise to the off-diagonal elements of Hamiltonian (\ref{Hamilton}).

One more important parameter describing the quantum dots is their spectral density distribution $n_{xE}$. We take it as a gaussian distribution over energy:
\begin{equation}
n_{xE}^i  = n_0 e^{ - \frac{{(E^i - E_{QD} )^2 }}{{\Delta E_{QD}^2 }}}
\label{gaussian}
\end{equation}
here $E_{QD}$ is the center of QDs distribution and $\Delta E_{QD}$ is its width.

\subsection{Equations for electromagnetic wave}

Dynamics of electromagnetic field in the microcavity in the presence of the pumped QDs is a rather complicated problem. In this paper we study modification of the lasing thresholds by the deformation pulse and are interested in not very strong deviation of the pump intensity  from the threshold one. Because of inevitable inhomogeneity of the QD spatial distribution there are always the preferential spots where the lasing starts first due to optimal gain conditions. These spots being of finite size feeds all cavity modes but are most efficient for the modes slowly propagating along the cavity i.e. the modes with in-plane wave vector close to zero\cite{Ramon_PRB}. Thus in what follows we will treat only normal cavity mode as the one that starts first to emit the coherent light. It should be understood that this assumption becomes invalid for the pump strongly ahead of the threshold value.

The dynamics of the electromagnetic field ${\cal E }_0$ of the normal cavity mode acting on the QDs can be found from the following oscillator-like equation:
\begin{equation}
i\hbar\frac{d}{dt}{\cal E}_0  = \hbar\omega_c {\cal E }_0  + \beta_J J_x
\label{waveeq}
\end{equation}
Here $\hbar\omega _c = Re(\hbar\omega _c) - i \gamma_c$ is the resonant photon energy of the microcavity for zero inplane wavevector. The imaginary part of cavity resonance
$\gamma _c$ results from the finite cavity lifetime. Second term $\beta_J J_x$ comes from the interaction with the resonant polarization  $J_x$ induced in QDs layer and coherent with the cavity electromagnetic field ${\cal E }_0$. This polarization can be written as
\begin{equation}
J_x  = \sum\limits_k {J_0^i n_{xE}^i \rho _{21}^i }
\label{Jx}
\end{equation}
For monochromatic wave with frequency $\omega$ we can solve Eq.(\ref{Neumannk}), find the $\rho_{21}^i$ and substitute it into Eq.(\ref{Jx}) to get:
\begin{equation}
J_x = {\cal E }_0
                    \sum\limits_k \frac{(J_0^i)^2}{\omega _0^i}
                    n_{xE}^i
                    \frac{1 - 2\rho _{22}^i}
                    {\hbar(\omega  - \omega_0^i)  - i\gamma _2}
\label{Jx2}
\end{equation}
In case of zero (or small) broadening of the QDs ensemble ($\omega_0^i=\omega_0$ the same for all quantum dots) this equation gives the standard polariton splitting equation for homogeneous exciton line.
\begin{equation}
 \hbar(\omega - \omega _c)
 \hbar(\omega  - \omega_0^i + i\gamma _2) =
 \beta_J
 \sum\limits_k \frac{(J_0^i)^2(1 - 2\rho _{22}^i)}{\omega _0}
                    n_{xE}^i
\label{omegac}
\end{equation}

For the opposite case of the broad QDs energy spectra that is the subject of our study we can calculate the modification of the imaginary part of the cavity resonance due to pumped QDs as
\begin{equation}
 g_c^{eff}  = Im[(\hbar\omega _c  + \beta_J \sum\limits_k {{\raise0.7ex\hbox{${(J_0^i )^2 }$} \!\mathord{\left/
 {\vphantom {{(J_0^i )^2 } {\omega _0 }}}\right.\kern-\nulldelimiterspace}
\!\lower0.7ex\hbox{${\omega _0 }$}}n_{xE}^i \frac{{1 - 2\rho _{22}^i }}{{\hbar\omega_c  - H_{22}^i - i\gamma _2 }})} ]
\label{omegac}
\end{equation}

\begin{figure}[h]
  \includegraphics[width=0.4\textwidth,clip]{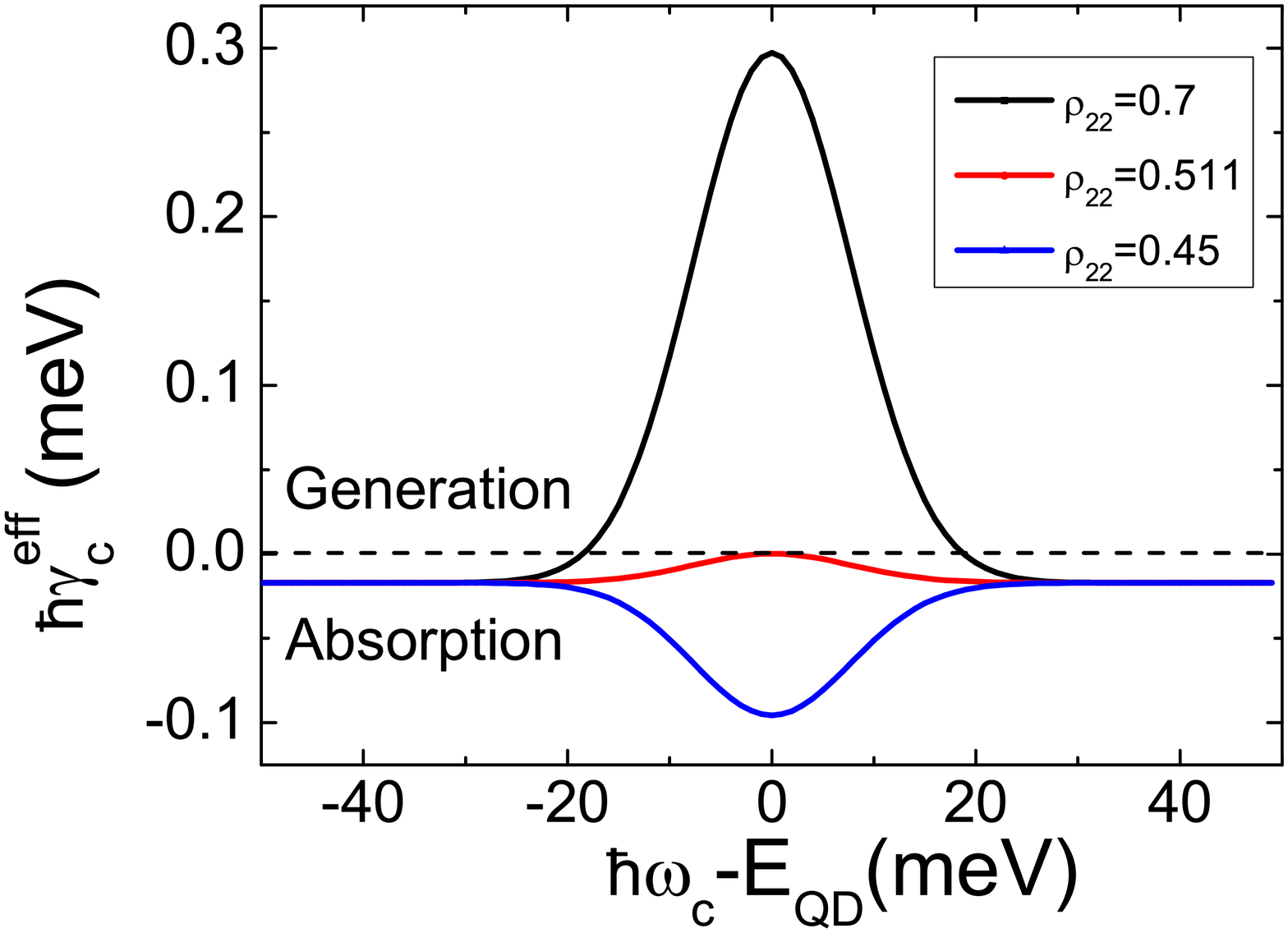}\\
  \caption{(Color online). Absorption coefficient dependence on detuning for different values of pumping.}
  \label{absorption}
\end{figure}

If $g_{c}^{eff} = -\gamma_{c}^{eff} < 0$ then the amplitude of electromagnetic field will decay with time $\hbar/\gamma_{c}^{eff}$ and the system is in absorption regime. In the cases when this parameter is positive the amplitude of the field grows and it means that the system is in generation regime. So, analyzing the $\gamma_{c}^{eff}$ we could analyze the generation-absorption transitions. On the fig. \ref{absorption} we've plotted the dependence of $\gamma_{c}^{eff}$ on detuning between cavity resonance and center of quantum dots distribution for three values of $\rho_{22}$.

From the picture it's clear that lasing threshold depends on pumping power as well as on detuning.

\subsection{Acousto-optical interaction}

To describe the interaction of quantum dots ensemble with acoustical vibrations we introduce a shift of the QD resonant energy proportional to the local deformation.

\begin{equation}
H_{22}^i (t)= E^i+E_{strain}(t)
\label{hamwithstrain}
\end{equation}
Here $E_{strain}(t)$ is the energy shift of exciton levels caused by the strain.

Now we have the equation describing the behavior and interactions between all three components: quantum dots, electromagnetic wave and acoustic vibrations.

\section{Implementation of the model}

\begin{figure}[h]
  \includegraphics[width=0.4\textwidth,clip]{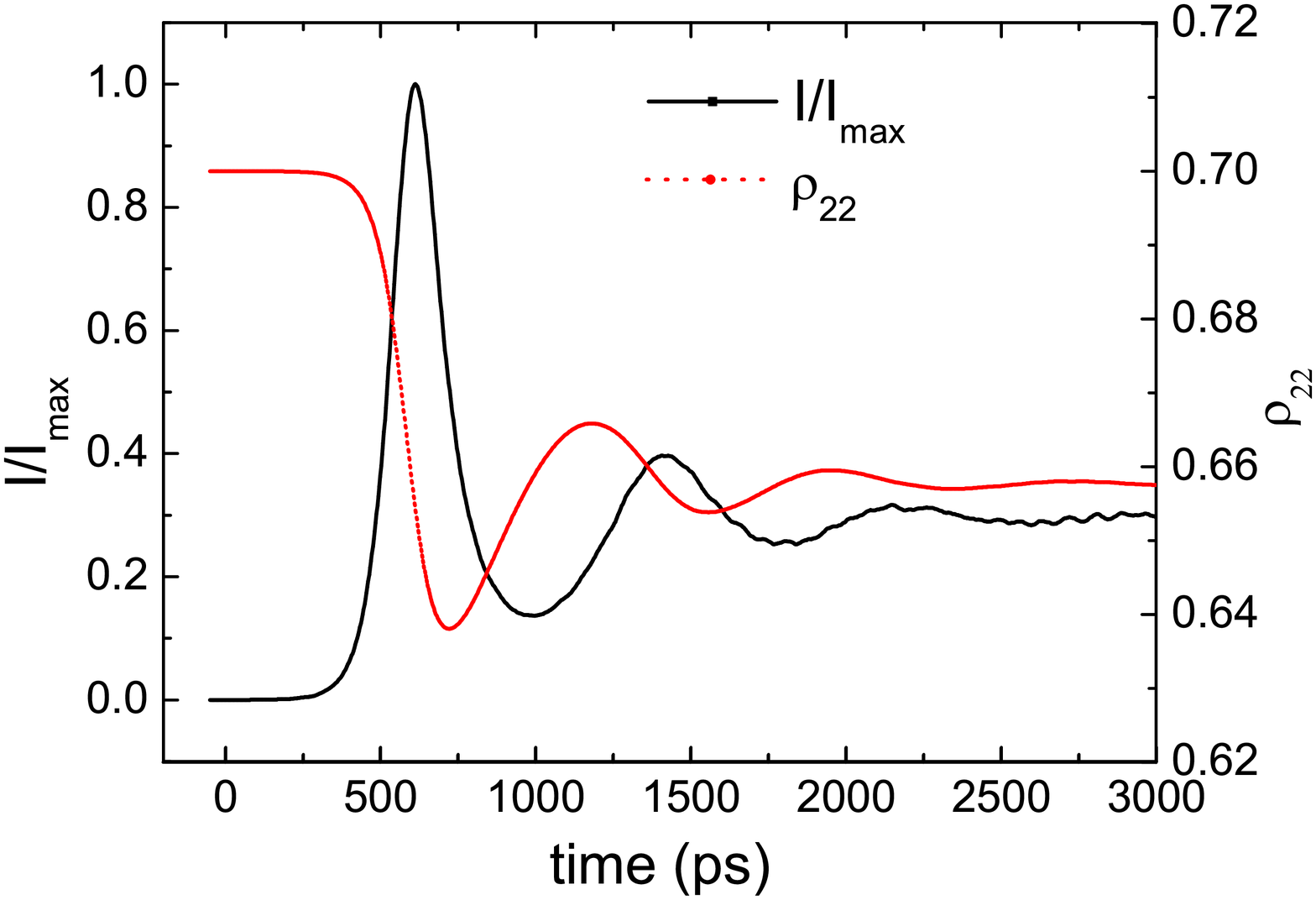}\\
  \caption{(Color online). Black curve - relation of photoluminescence intensity to its maximum value versus time. Red dotted curve - time evolution of $\rho_{22}$}
  \label{Eqwtime}
\end{figure}

Let's first consider the time-evolution of our model system without any strain pulse
We take $\rho^i_{22}=\rho^{st}_{22}=0.7$ at starting point and we consider the detuning
$\delta=\hbar \omega_c - E_{QD}=12 meV$ which corresponds to slightly above lasing threshold regime. Results of the calculations are shown in Fig. \ref{Eqwtime}. The black curve shows the time evolution of the photoluminescence intensity, and the red dotted curve shows $\rho_{22}$ for quantum dots that are in the resonance with the microcavity. One can see that both electric field and the occupation $\rho_{22}$ oscillate in time with the same frequency but shifted in a phase by $\frac{\pi}{4}$. These oscillations show periodical transitions of the system from lasing to absorption regime and vice versa.
This effect comes from assumption of systems homogeneity and for realistic non-homogeneous cases we should average in time the intensity coming from different lasing spots.

\begin{figure}[h]
  \includegraphics[width=0.4\textwidth,clip]{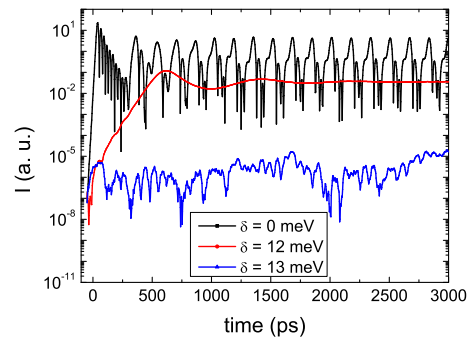}\\
  \caption{(Color online). Intensities of electromagnetic fields calculated for different values of detunings.}
  \label{Eqwtime2}
\end{figure}

Typical time dependencies of the electromagnetic field intensities for different detunings $\delta=\hbar \omega_c - E_{QD}$ are shown on Fig. \ref{Eqwtime2}. The black curve shows the case when the cavity resonance coincides with the QDs distribution maximum. The red and blue ones are for the detunings slightly above and slightly below lasing threshold respectively. Electromagnetic radiation in the absorption regime is defined by spontaneous radiation only and it is a narrow-band noise with small amplitude.

\begin{figure}[h]
  \includegraphics[width=0.4\textwidth,clip]{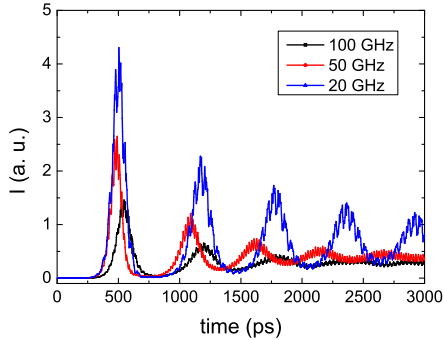}\\
  \caption{(Color online). Intensities of electromagnetic fields calculated for different frequencies of harmonic acoustic vibrations.}
  \label{diffreq}
\end{figure}

The application of the acoustic pulse change drastically the dynamics of the microcavity.
First we took acoustic vibrations in a form of model harmonic oscillations. We put the detuning $\delta=\hbar \omega_c - E_{QD}=12 meV$, that corresponds to slightly above lasing threshold case. The amplitude of vibrations we took equal to 1 meV and we calculated several curves for different frequencies of oscillations. The results for 100, 50 and 20 GHz are shown on Fig. \ref{diffreq}. From this picture one can see that even small vibrations can strongly amplify the signal from microcavity. This effect comes from the fact that in the presence of vibrations in the system, more quantum dots are involved in lasing process. The characteristic time of microcavity light coupling to the resonant QDs is proportional to the dots density. The amplification of cavity field depends on vibration frequency. When these vibrations are too frequent, the quantum dots just don't have time to respond on external field.

\begin{figure}[h]
  \includegraphics[width=0.4\textwidth,clip]{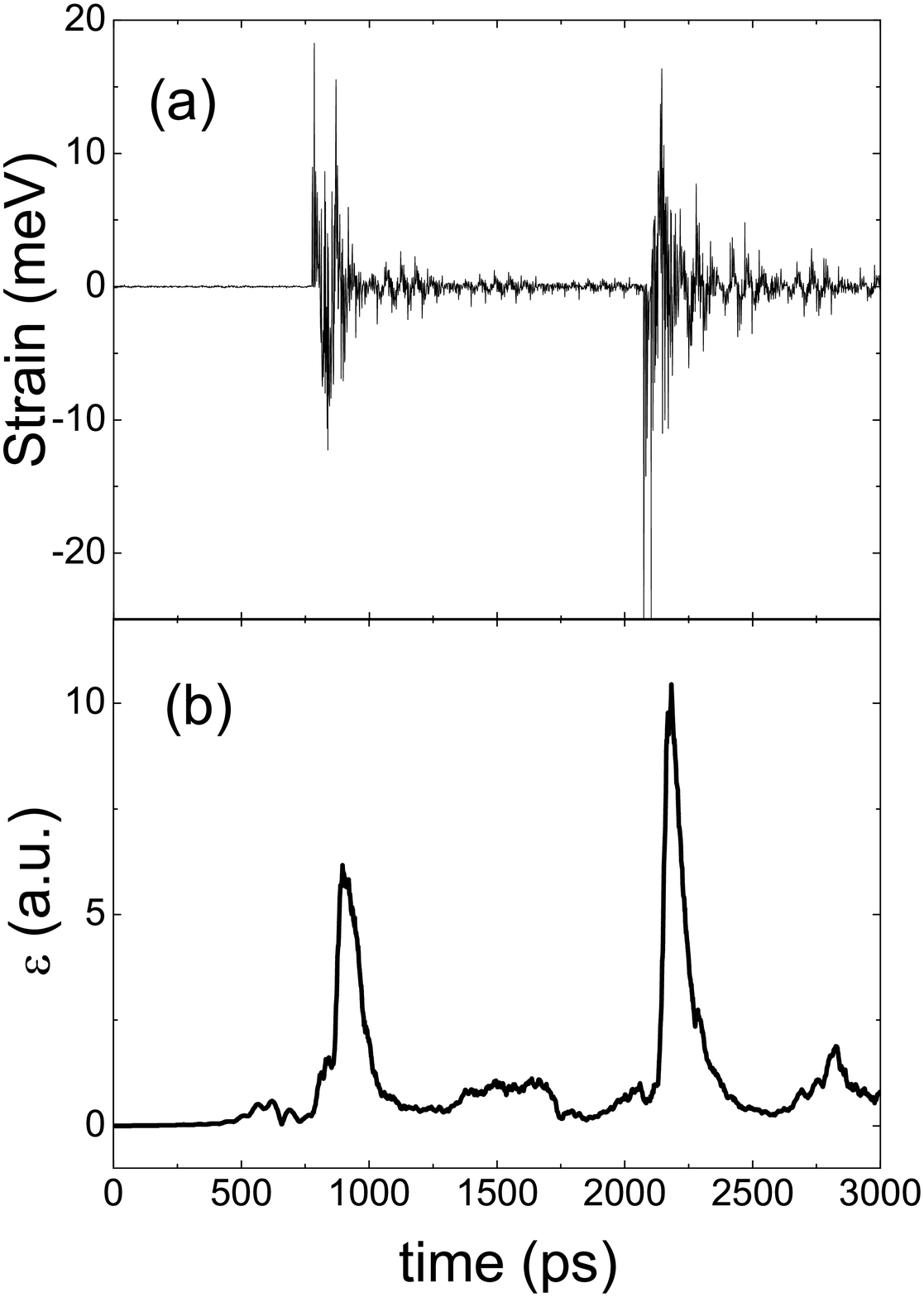}\\
  \caption{(a) Temporal profile of real strain. (b) Evolution of the electromagnetic fields amplitude in MC.}
  \label{realstrain}
\end{figure}

In order to compare the results of our approach with the experimental work \cite{Bruggemann} we took their strain pulses and put it into our model.
The profile of the strain pulse is shown in Fig. \ref{realstrain} (a).
Incident strain pulse comes in the moment of approximately 750 ps. Its starts from compressive deformation that increases the gap and the energy of QDs excitons.
At the moment of 2100 ps the pulse reflected from the surface comes which starts with decompression part. In the experimental work \cite{Bruggemann} authors surprisingly observed that amplification of luminescence intensity from the reflected pulse is much stronger that from the incident one. To check whether it's possible to obtain this effect within our model, we put the real strain profile in our code and calculated the time evolution of the electromagnetic field amplitude. One can see it on fig. \ref{realstrain} (b). And this effect was successfully reproduced.

We propose the following  explanation of the physics of this effect. One can notice that between two pulses there still remain small vibrations in the system. Their amplitude is not larger than 1 meV but as we have shown in the beginning of this section even such small oscillations could strongly modify electromagnetic field in cavity. From Fig. \ref{realstrain} (b) it is clear that between pulses the amplitude of the field is much larger than before the incident pulse. So the initial conditions are "better" for amplification before the second pulse than before the first one when the amplification factor of the reflected pulse could be less than of the incident.

\begin{figure}[h]
  \includegraphics[width=0.4\textwidth,clip]{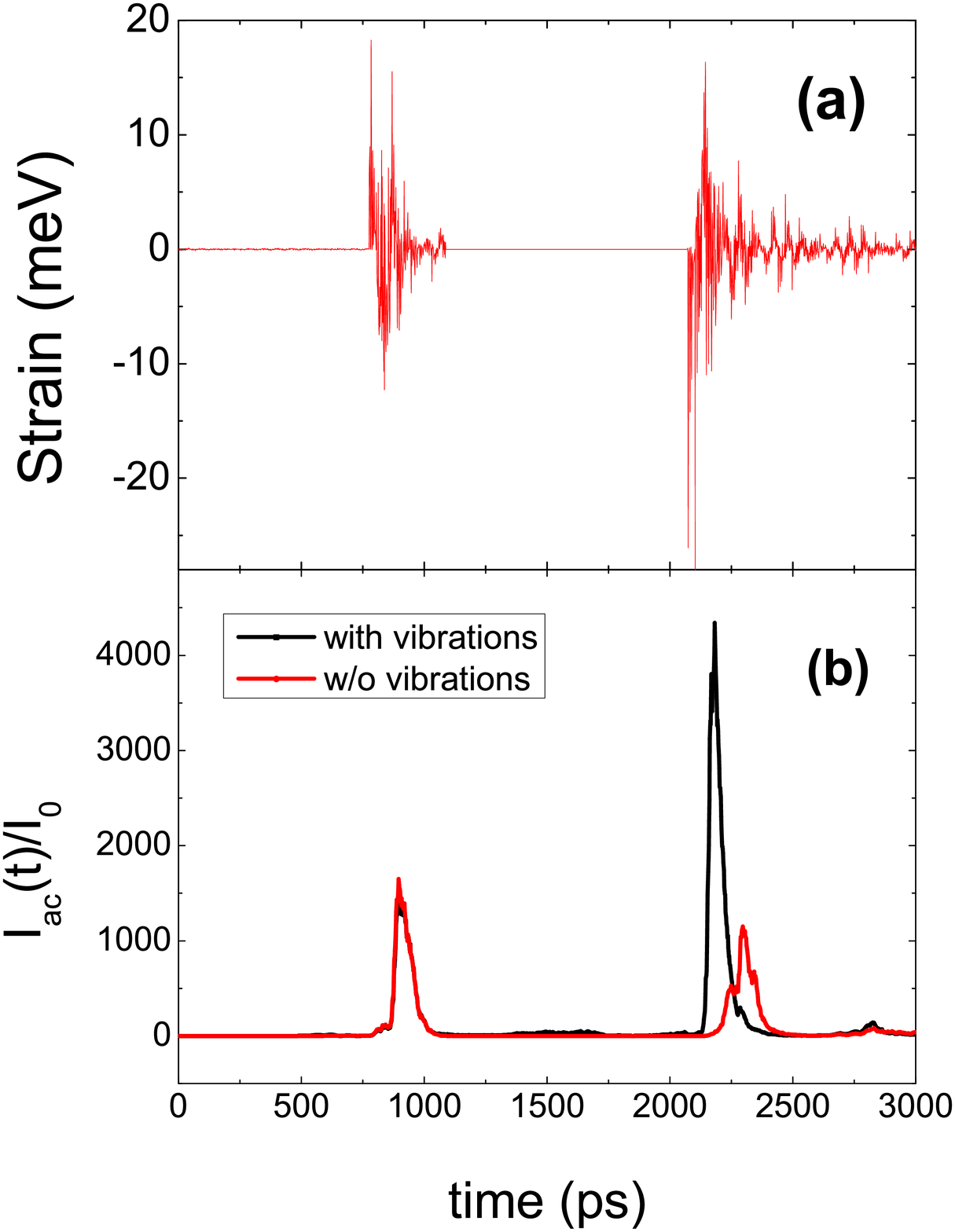}\\
  \caption{(Color online) (a) Temporal profile of real strain without interpulse vibrations. (b) Amplification of the PL intensities with (black curve) and without (red curve) vibrations.}
  \label{realstrain2}
\end{figure}

To check our assumptions we cut the vibrations between two pulses and put the new strain profile in our model. One can see it in Fig. \ref{realstrain2} (a). In the figure  \ref{realstrain2} (b) we've shown the ratio between the intensity in the presence of acoustic pulses ($I_{ac}(t)$) and the average intensity without them ($I_{0}$) for two cases: with (black curve) and without (red curve) the interpulse vibrations.

It is clear that the real amplification from the reflected strain pulse is less than from the incident.

\section{Modulation of the lasing by surface standing acoustic waves}

Surface acoustic waves (SAW) are interesting objects to study light-matter coupling in semiconductor systems. Because of its long wavelengths they produce slowly changing in time (quasistationary for excitons in most planar systems) periodic potential. Recently SAW were used in studies of polariton condensates \cite{PRL_SAW_polariton, NJP_SAW_polariton, Vishnevsky}, photoluminescence from quantum wells \cite{PRB_SAW_QW} etc. We propose to implement SAW to obtain different lasing patterns in the ensemble of quantum dots coupled to microcavity. If we will consider two counterpropagating in xy-plane acoustic waves with the same amplitude we could obtain standing SAW (SSAW). Furthermore we can produce two-dimensional SSAW by the interference between two orthogonaly propagating one-dimensional SSAW. The $E_{strain}$ in this case will be the function of planar coordinates and time and can be written in the form:

\begin{equation}
E_{strain}(x,y,t)= A_{SAW}\sin{(k_{x}x)}\sin{(k_{y}y)}\sin{(\omega_{SAW}t)}
\label{sawstrain}
\end{equation}
Here $A_{SAW}$ - amplitude of exciton energy shifting by SAW (several meV in order), $k_{x}, k_{y}$ - wave-vectors of $x-$ and $y-$SAW, $\omega_{SAW}$ - frequency of SAW.

It was shown in Sec. IIIa that for the same pump intensity, transition to the lasing regime strongly depends on the relative position between the cavity resonance and quantum dots distribution maximum. So for given pump power there is threshold value of detuning $\delta_{th}$ that when $|\delta| \le \delta_{th}$ there is a gain in the system. And it is clear that using strain provided by planar acoustic waves we can obtain some regions of cavity where conditions of lasing would be satisfied and in others would not. So we could observe patterns of bright spots. To derive the dependence of lasing intensity on planar coordinates we've considered a simple model based on a pair of kinetic equation:
\begin{equation}
\frac{{dN_{ph} }}{{dt}} = wN_x (2\rho _{22}  - 1)N_{ph}  - \frac{{N_{ph} }}{{\tau _c }}
\label{dnph}
\end{equation}
\begin{equation}
N_x \frac{{d\rho _{22} }}{{dt}} = P - wN_x (2\rho _{22}  - 1)N_{ph}  - \frac{{N_x \rho _{22} }}{{\tau _{QD} }}
\label{dnx}
\end{equation}
\begin{widetext}
\begin{figure}[h]
  \includegraphics[width=1\textwidth,clip]{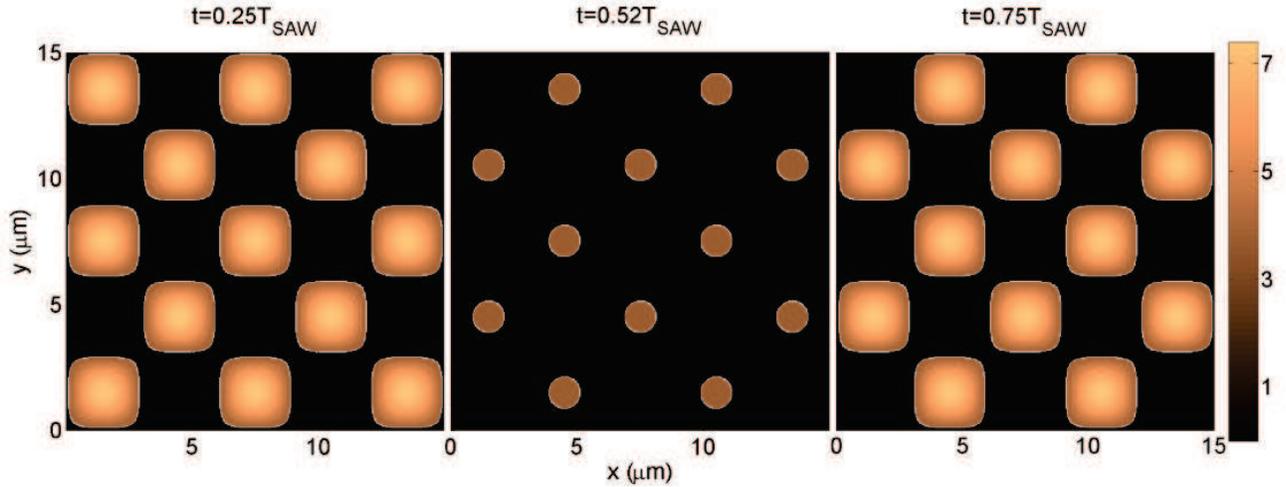}\\
  \caption{(Color online) Evolution of photoluminescence in time. Snapshots are taken for the values of time 0.25, 0.52 and 0.75 of the SAW period $T_{SAW}$}
  \label{SAWfig}
\end{figure}
\end{widetext}

 First equation describes number of photons ($N_{ph}$) in system and second one - number of excited quantum dots ($N_x\rho_{22}$). $w$ - is a probability for the photon to be captured by free quantum dot, $\tau_c$ - lifetime of photon in MC when $\tau_{QD}$ is a non-radiative lifetime of an excited quantum dot. $P$ is a term describing the pump and in general case it should be proportional to the number of free quantum dots: $P=pN_{x}(1-\rho_{22})$. Here we consider only quantum dots which take participation in lasing, in other words which frequency of transition is equal to the MC frequency:

\begin{equation}
N_x(x,y,t) \sim n_0 exp( - \frac{{(E_{mc} - E_{QD} +E_{strain}(x,y,t))^2 }}{{\Delta E_{QD}^2 }})
\label{Nxgaus}
\end{equation}

In the regions of absorption regime we can take $N_{ph}=0$. In other regions we could derive $N_{ph}$ finding stationary solutions of Eqs.(\ref{dnph}-\ref{dnx}):

\begin{equation}
N_{ph}  = \tau _c \frac{{p\tau _{QD}  - 1}}{{2\tau _{QD} }}N_x  - \frac{{p\tau _{QD}  + 1}}{{2w\tau _{QD} }}
\label{Nphstat}
\end{equation}

From this equation conditions for the lasing could be derived. First - pump should be strong enough so the prefactor before $N_x$ to be positive: $p\tau _{QD}\ge{1}$. Second, even with strong pumping it is necessary to have enough quantum dots to make $N_{ph}$ positive, so $\tau _c (\frac{{p\tau _{QD}  - 1}}{{2\tau _{QD} }})N_x  \ge \frac{{p\tau _{QD}  + 1}}{{2w\tau _{QD} }}$.

On Fig. \ref{SAWfig} we plotted the solutions of Eqs. (\ref{Nxgaus}-\ref{Nphstat}) for different moments of time. We consider the pump corresponding to the threshold detuning $\delta_{th} = 10 meV$, and the unstrained detuning of system we took to be slightly more. $A_{SAW}$ we took equal to $5 meV$. The video of time evolution of photoluminescence could be seen at \cite{video1}.

\begin{figure}[h]
  \includegraphics[width=0.5\textwidth,clip]{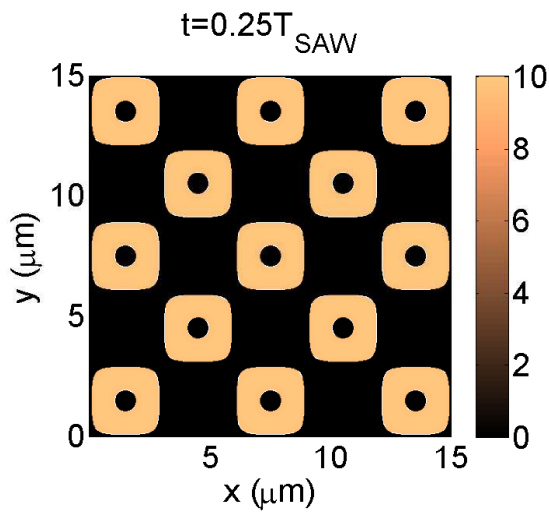}\\
  \caption{(Color online) Snapshop of PL for $\delta \approx \delta_{th}=2 meV$ and $A_{SAW}=5 meV$}
  \label{cirsaw}
\end{figure}

Now let's consider weaker pumping to obtain threshold detuning $\delta_{th}$ less than double strain amplitude. If we start with absolute value of detuning $\delta$ slightly larger than $\delta_{th}$ then the strain could be so high that $|\delta|$  will cross the threshold twice. In this case one can obtain holes in bright spots of photoluminescence as it is shown on fig. \ref{cirsaw}. Video of this process could be seen at \cite{video2}.

\section{Summary}

In our work we've developed theoretic model of light-matter interaction in the system of quantum dots coupled to microcavity in the presence of acoustic deformations. We've described the effect of acoustically driven amplification of the lasing regime in the system. Moreover, we successfully reproduced experimental effects obtained in \cite{Bruggemann}. At the final section of our work we proposed an implementation of surface acoustic waves to modify lasing patterns of the system.

\section*{Acknowlegements}

This work was supported by EU ITNs "Spin-Optronics" Grant No. 237252 and INDEX Grant No. 289968. Also we would like to thank C. Bruggemann, A. V. Scherbakov, A. V. Akimov, D. R. Yakovlev, M. Bayer, I. V. Ignatiev, V.D. Kulakovskii and G. Mapluech for useful discussions.

\end{document}